\documentclass[aps,twocolumn]{revtex4-1}
\usepackage{graphicx}
\usepackage{epstopdf}
\usepackage{amsmath}
\usepackage{multirow}

\begin{document}

\title{First-principles study of FeSe epitaxial films on SrTiO$_3$}

\author{Kai Liu$^{1}$}
\author{Miao Gao$^{2}$}
\author{Zhong-Yi Lu$^{1}$}\email{zlu@ruc.edu.cn}
\author{Tao Xiang$^{2}$}\email{txiang@iphy.ac.cn}

\affiliation{$^{1}$Department of Physics and Beijing Key Laboratory of Opto-electronic Functional Materials $\&$ Micro-nano Devices, Renmin University of China, Beijing 100872, China}

\affiliation{$^{2}$Institute of Physics, Chinese Academy of Sciences, Beijing 100190, China}

\date{\today}

\begin{abstract}
The discovery of high temperature superconductivity in FeSe films on SrTiO$_3$ substrate has inspired great experimental and theoretical interests. First-principles density functional theory calculations, which have played an important role in the study of bulk iron-based superconductors, also participate in the investigation of interfacial superconductivity. In this article, we review the calculation results on the electronic and magnetic structures of FeSe epitaxial films, emphasizing on the interplay between different degrees of freedom, such as charge, spin, and lattice vibrations. Furthermore, the comparison between FeSe monolayer and bilayer films on SrTiO$_3$ is discussed.
\end{abstract}


\maketitle

\section{Introduction} 

Iron-based superconductors\cite{1111, 122, 111, 11, luo} are another class of materials with superconducting transition temperature $T_c$ exceeding 40 K in addition to cuprates. They all contain the conducting \textit{X}-Fe$_2$-\textit{X} (\textit{X}=As, Se, Te \textit{et al.}) layers and most of their parent compounds are antiferromagnetic (AFM) semimetals.\cite{MaFJ08-2} Their superconducting transition temperature $T_c$ can be altered by chemical doping or by applying high pressure. It is commonly believed that the superconducting pairing occurs in the \textit{X}-Fe$_2$-\textit{X} layer, and the superconducting temperature is strongly affected by antiferromagnetic fluctuations.

Epitaxial film growth provides a new route to tune $T_c$ for Fe-based superconductors.\cite{CaoLX, SongCL, WangQY, FengDL-KTO, FengDL-BTO} In 2012, atomically flat FeSe films under tensile strain were grown on SrTiO$_3$ with molecular-beam epitaxy technique.\cite{WangQY} The monolayer FeSe shows an even higher superconducting transition temperature than bulk FeSe under ambient pressure ($\sim$9 K). However, the double-layer FeSe film on the SrTiO$_3$ substrate does not superconduct and its low-temperature scanning tunneling microscope (STM) spectrum shows a tiny semiconducting gap.\cite{WangQY} From the angle-resolved photoemission spectroscopy (ARPES) measurement, it was found that the FeSe monolayer on SrTiO$_3$ contains only electron-type Fermi surfaces located at the corners of Brillouin zone,\cite{ZhouXJ12, ZhouXJ13, FengDL13} unlike in the bulk FeSe or in iron-pnictide superconductors. A number of studies suggested that charge transfer from SrTiO$_3$ to the FeSe layer, introduced either by Nb doping in SrTiO$_3$ or by oxygen vacancies during annealing, has a strong effect on the superconducting pairing state,\cite{XueQKAPE13, XueQKPRB14, MaXC14, ZhouXJPNAS14, ZhouXJNC14} and the superconductivity emerges only when the antiferromangetic correlation is strongly suppressed.\cite{FengDL13} In addition to this, the strain effect has also been studied. It was found that $T_c$ could be enlarged by using substrates with larger lattice constants.\cite{FengDL-KTO, FengDL-BTO}  A replica band, which is about 100 meV below the main band around $M$, was observed by the ARPES measurement on the monolayer FeSe.\cite{ShenZX14} This replica band was suggested to result from the coupling between FeSe electrons and substrate phonons, which may strengthen the pairing force and enhance $T_c$.\cite{ShenZX14}

The first-principles density functional theory calculations have played an important role in the study of the physical properties of FeSe epitaxial films. The band structures and the leading magnetic instability for the single- and double-layer FeSe films on SrTiO$_3$ substrates were first investigated by Liu \textit{et al.}\cite{LiuK12,LiuK15} More systematical theoretical investigations were done by a number of other groups to reveal the role of spin, charge, orbital, lattice, and other degrees of freedoms, the interplay among these degrees of freedoms, and the magnetic configurations of the ground states.\cite{XiangYY12, Bazhirov13, ZhangP13, ZhangSB13, KuW14, GongXG14, GongXG15, XingDY14, Coh14, HuJP14, Profeta15, GongXG15b}

The first-principles calculations were nearly all carried out using the ultrasoft pseudopotential (PP) method,\cite{Vanderbilt90} as implemented in the Quantum-Espresso,\cite{Giannozzi09} or the projector augmented wave (PAW) method,\cite{Blochl94, Kresse99} as in the VASP package.\cite{Kresse93, Kresse96a, Kresse96b} The generalized gradient approximation (GGA) of Perdew-Burke-Ernzerhof\cite{Perdew96} for the exchange-correlation potentials was often adopted. An LDA+U calculation was done by Zheng \textit{et al.}\cite{ZhangP13} to simulate approximately the correlation effect of Fe 3$d$-electrons. Both FeSe film and SrTiO$_3$ substrate are layered materials. The van der Waals (vdW) interaction plays an important role in the interlayer bonding between any two neighboring layers. This interaction was generally ignored in the density functional theory calculations, but was shown to be important for the FeSe epitaxial films on SrTiO$_3$ by Liu \textit{et al.}\cite{LiuK15}

The rest of the article is organized as follows. In Section 2, we will discuss the charge doping effect. Studies on electron-phonon coupling will be reviewed in Section 3. In Section 4, we will discuss the magnetic orders that have been calculated, and in Section 5, the spin-phonon coupling will be the focus. The comparison between FeSe monolayer and bilayer films will be given in Section 6. Finally, the article will end with a summary and perspective in Section 7.


\section{Charge doping effect}

The monolayer FeSe on the SrTiO$_3$ substrate shows a strong charge effect. The as-grown FeSe film is generally non-superconducting. It becomes superconducting after annealing, which induces doping of electrons by introducing oxygen vacancies in the substrate. As revealed by the ARPES measurements, the superconducting monolayer FeSe film contains two electron-like Fermi surfaces around the corners of the Brillouin zone,\cite{ZhouXJ12, ZhouXJ13, FengDL13} unlike that in the bulk FeSe which has also a hole-like Fermi surface at the Brillouin zone center. The absence of hole-like Fermi surface in the single-layer FeSe film indicates that the nesting effect between electron- and hole-like Fermi surfaces is not important, at least in this material.

Soon after the discovery of superconductivity in the FeSe epitaxial films,\cite{WangQY} we evaluated the adsorption energies for the single and double FeSe layers on both TiO$_2$-terminated and SrO-terminated SrTiO$_3$ surfaces.\cite{LiuK12} We found that the TiO$_2$-terminated surface with Se atoms located right above Ti atoms is the most stable lattice configuration. If the doping effect from Nb dopants or oxygen vacancies is absent in the substrate, there is only a charge redistribution at the interface, but without charge transfer between FeSe and SrTiO$_3$.\cite{LiuK12, GongXG14}



In the density functional theory calculation, the charging effect on the single-layer FeSe film was mimicked either by doping electrons\cite{Bazhirov13, ZhangP13, LiuK15} or by introducing anion vacancies, such as oxygen vacancies.\cite{ZhangSB13, KuW14, GongXG14} It was found that both the density of states at the Fermi level\cite{Bazhirov13} and the band dispersion around the Brillouin zone center depend strongly on the doping level for the free-standing FeSe monolayer.\cite{ZhangP13} Oxygen vacancies in SrTiO$_3$ also play an important role in adhering the FeSe film to the substrate.\cite{ZhangSB13} Berlijn \textit{et al.} studied the effect of Se vacancies and found that Se vacancies behave like hole dopants rather than electron dopants in the monolayer FeSe.\cite{KuW14} This rules out the possibility of Se vacancies as the origin of the large electron pockets observed in the ARPES measurements. From the spatial distribution of charge carriers, we found that doped electrons are mainly distributed at the interface.\cite{LiuK15} Meanwhile, the adsorption of FeSe layer on the substrate induces a charge redistribution around Fe atoms.\cite{LiuK15}




\section{Electron-phonon coupling}

The pairing mechanism in the single-layer FeSe film is still unknown. But it is generally believed that the conducting electrons in the FeSe films couple strongly with phonon modes in the SrTiO$_3$ substrate. Based on a theoretical investigation on the screening effect of ferroelectric phonons of SrTiO$_3$ on the pairing interaction in the FeSe layer, Xiang \textit{et al.} suggested that the superconducting transition driven by spin fluctuations can be strongly enhanced by the electron-phonon coupling, which in turn can increase significantly the superconducting transition temperature and even alter the pairing symmetry.\cite{XiangYY12} This picture of strong electron-phonon coupling was supported by the observation of a replica band, which is about 100 meV below the main band, in the ARPES measurement for the single-layer FeSe/SrTiO$_3$ film.\cite{ShenZX14} The observed replica band is believed to result from the coupling of low-energy electrons with oxygen optical phonon modes of $\sim$80 meV generated by oxygen vacancies in SrTiO$_3$.\cite{GongXG15b} In the FeSe/SrTiO$_3$ film capped with FeTe layers, an exotic acoustic phonon mode, which seems to couple destructively with the pairing interaction, was also observed in the recent ultrafast spectroscopy measurement.\cite{ZhaoJM15}

Li \textit{et al.} carried out a first-principles calculation for the electron-phonon coupling constant of the single-layer FeSe/SrTiO$_3$ film.\cite{XingDY14} They found that the electron-phonon coupling in this material arises mainly from the interaction between the interfacial ferroelectric phonons (with frequencies 390, 450, and 690 cm$^{-1}$) and electrons in the FeSe layer. By utilizing the strong coupling theory of superconductivity, they found that the superconducting transition induced by the electron-phonon coupling is about 2.6 K, higher than the corresponding value for the bulk FeSe, but still one order of magnitude smaller than the experimental value.\cite{XingDY14} By adjusting empirically the interacting potential to reproduce the electronic band structure as observed by the experiment, Coh \textit{et al.} found that the electron-phonon coupling can be greatly enhanced and the superconducting transition temperature can be as high as 38 K for the single-layer FeSe film in an antiferromagnetic state if the in-plane lattice constants take the same values as for SrTiO$_3$(001).\cite{Coh14} There are two phonon modes that couple strongly with electrons. Their frequencies are equal to 10 meV and 20 meV, respectively.



\section{Magnetic order}

The interplay between superconductivity and magnetic order is important to the understanding of pairing mechanism in iron-based superconductors. In bulk FeSe, the long-range magnetic order was not observed at ambient pressure down to low temperature, but the antiferromagnetic fluctuation, as revealed by the nuclear magnetic resonance experiment, is strong.\cite{Cava09} By applying a pressure of about 1 GPa, a static antiferromagnetic order coexisting microscopically with superconductivity can be stabilized in this material. This is confirmed by the muon-spin rotation and relaxation ($\mu$SR) and magnetization measurements.\cite{Bendele10} For the FeSe epitaxial film on SrTiO$_3$, it is difficult to measure its magnetic order. But a band splitting associated with an antiferromagnetic order or instability was observed in this material by the ARPES measurements.\cite{FengDL13} This kind of splitting was also observed in BaFe$_2$As$_2$ and many other parent compounds of iron pnictides whose ground states are in collinear antiferromagnetic order.\cite{FengDL09}

\begin{figure}
\begin{center}
\includegraphics[scale=0.2]{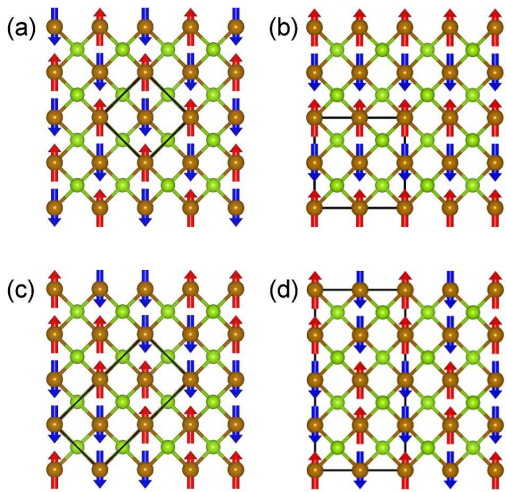}\\[5pt]  
\end{center}
\caption{(color online) Possible magnetic orders in FeSe: (a) checkerboard AFM N\'eel order, (b) collinear AFM order, (c) bicollinear AFM order, and (d) pair-checkerboard AFM order. The red and blue arrows represent the up and down spins of Fe ions, respectively. }\label{fig:1}
\end{figure}

Figure \ref{fig:1} shows several possible magnetic orders for FeSe/SrTiO$_3$ films. Early calculations concentrated on the states with ferromagnetic order, checkerboard AFM N\'eel order, collinear AFM (single-stripe) order, and bicollinear AFM (double-stripe) order.\cite{MaFJ08, MaFJ09} Among these states, it was generally found that the energy of the collinear AFM order is the lowest.\cite{LiuK12, LiuK15, Bazhirov13, GongXG14} However, recently, it was found that the lowest energy state is the pair-checkerboard AFM state whose magnetic order is shown in Fig. \ref{fig:1}(d).\cite{GongXG15} Furthermore, Cao \textit{et al.} found that the substrate-induced tensile strain tends to stabilize the collinear AFM order in the FeSe film, but the charge transfer induced by oxygen vacancies in the substrate will suppress the magnetic order.\cite{GongXG14}

\begin{figure}
\begin{center}
\includegraphics[scale=0.3]{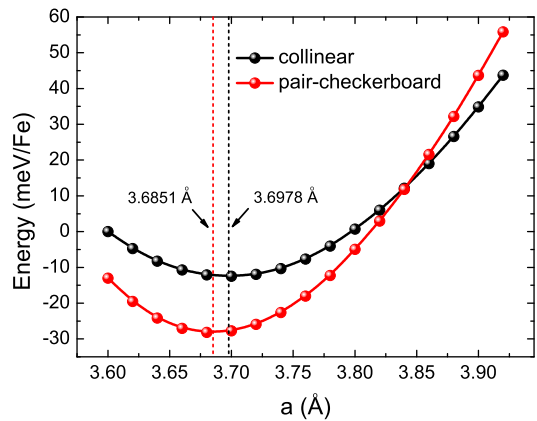}\\[5pt]  
\end{center}
\caption{(color online) Total energies of collinear and pair-checkerboard AFM orders versus the in-plane lattice constant for bulk FeSe. The unit cells for these two AFM ordered states are $\sqrt{2}a\times\sqrt{2}a\times c$ and $\sqrt{2}a\times2\sqrt{2}a\times c$, respectively. The optimal in-plane lattice constants are marked by the black and red dashed lines for the collinear and pair-checkerboard AFM states, respectively.}
\label{fig:2}
\end{figure}

In order to further understand the effect of in-plane tensile strain on FeSe layers, we evaluated the total energy as a function of the in-plane lattice constant for the bulk FeSe in both the collinear and the pair-checkerboard AFM states and the result is shown in
Fig. \ref{fig:2}. In the calculations, the van der Waals correction is included and the lattice constant along the $c$-axis with the internal atomic positions are relaxed to minimize the total energy. The optimized pair-checkerboard AFM state is found to be 15.6 meV/Fe lower in energy than the corresponding collinear AFM state, in agreement with the previous calculation.\cite{GongXG15} But the energy gap between these two states decreases with the increase of the in-plane lattice constant. The energy of the collinear AFM state is lower than the pair-checkerboard AFM state when the in-plane lattice constant becomes larger than 3.85 {\AA}. A similar case happens for the freestanding monolayer FeSe film, as pointed out by Tresca \textit{et al.}\cite{Profeta15} The strong competition between the collinear and pair-checkerboard AFM ordered states suggests that there are strong magnetic frustrations in both bulk and monolayer FeSe. As no long-range AFM order is observed in the bulk FeSe, this could be a strong indication that FeSe has a quantum spin liquid-like state, which is intimately related to the fascinating high-$T_c$ superconductivity in FeSe-based materials.


\section{Spin-phonon coupling}


For the bulk FeSe the transition temperature can arise from 8.3 K at ambient pressure to ~37 K under high pressure,\cite{Felser09, Mizuguchi10, Margadonna09} accompanied by a dramatic reduction of the lattice constant along the $c$ axis.\cite{Felser09} For the FeSe epitaxial film, its superconducting transition temperature is also known to depend strongly on the lattice constants of the substrates.\cite{FengDL-KTO, FengDL-BTO} This suggests that the spin-phonon coupling is strong in these materials since the pairing interaction is closely correlated with the antiferromagnetic fluctuation.


Using the first-principles electronic structure calculations by including the van der  Waals correction, we determined the lattice structures of bulk FeSe under pressure as well as FeSe films on SrTiO$_3$ under different electron doping. We found that both the lattice constants and the phonon frequencies for bulk FeSe at ambient pressure agree well with the experimental values.\cite{YeQQ13} The frequencies of all optical phonon modes at the Brillouin zone center increase with pressure. Among them, the $A_{1g}$ mode, which consists of coherent vertical vibrations of Se atoms relative to the Fe-Fe plane, shows an abrupt frequency jump between 5 GPa and 6 GPa. The magnetic moment is strongly affected by this mode. The atomic displacements due to the zero-point vibrations of different optical phonons can induce the variations of local magnetic moment on Fe and change the ordering moment under pressure. This is consistent with the experimental observation that AFM spin fluctuations in FeSe are enhanced by  pressure.\cite{Cava09} It suggests that the effect of phonon through spin-phonon coupling is important and should not be ignored.



For FeSe epitaxial films on SrTiO$_3$, the induced variation of magnetic moment by the zero-point atomic displacement of the phonon mode is less than 0.2 $\mu_B$,\cite{LiuK15} much smaller than that ($\sim$ 1.0 $\mu_B$) in the bulk FeSe.\cite{YeQQ13} The change is mainly induced by the $A_1$ mode, whose atomic displacement is similar to the $A_{1g}$ mode of bulk FeSe.
The energy difference between the collinear AFM state and the checkerboard AFM N\'eel state is almost unchanged by the zero-point atomic displacement of vertical phonon modes for the undoped FeSe epitaxial film, but strongly changes upon 0.2 electron doping.\cite{LiuK15}
Magnetic frustration would become stronger if the energy difference between different magnetic states is reduced.

\section{Comparison between FeSe monolayer and bilayer films}

From the STM spectra of FeSe/SrTiO$_3$, it is known that the monolayer FeSe film behaves differently from the double-layer FeSe film. The monolayer FeSe upon electron doping shows a superconducting gap while the bilayer FeSe does not.\cite{WangQY} However, from the first-principles calculations, we found that the electronic band structures of monolayer and bilayer FeSe are not much different.\cite{LiuK12} In particular, we found that for the monolayer FeSe, the energy of the checkerboard AFM N\'eel state is always lower than that of the nonmagnetic state and the energy difference between these two states varies very little with doping (within 2 meV/Fe). However, the energy difference between the collinear AFM state and the checkerboard AFM N\'eel state reduces considerably with electron doping, and remains almost unchanged when more than 0.2 electrons are doped. Thus the electron doping has a notable effect on the magnetic properties of monolayer FeSe.\cite{LiuK15}


For the bilayer FeSe, the checkerboard AFM N\'eel state has a lower energy than the nonmagnetic state. But the energy difference between the collinear AFM and the checkerboard AFM N\'eel states reduces with electron doping. In order to reduce the energy difference, for example, to -17 meV per Fe, the doping concentration is significantly larger in the double-layer system than that in the single-layer one.\cite{LiuK15} This agrees with the recent experimental observation that the bilayer FeSe epitaxial film is more difficult to be doped than the monolayer FeSe.\cite{ZhouXJNC14} Moreover, the bilayer FeSe can have different kinds of magnetic orders on the bottom and top layers. By heavily increasing the doping level, the state with the bottom layer in the checkerboard AFM N\'eel order and the top layer in the collinear AFM order can become the ground state.\cite{LiuK15}


\section{Summary and perspectives}

We have briefly reviewed the first-principles density functional theory calculations on the electronic and magnetic structures of FeSe/SrTiO$_3$ films and compared with the corresponding results for the bulk FeSe. Without doping, the ground state of FeSe epitaxial films was found to be antiferromagnetically ordered. However, the energy difference between different antiferromagnetically ordered states is very small. This may indicate that the magnetic states in these materials are highly frustrated, which may eventually melt the antiferromagnetic order, leading to a quantum spin liquid state, as observed by experiments. Upon doping, the electronic band structure is changed and the density of states at the Fermi level is increased. Furthermore, the electron-phonon coupling is enhanced at the interfacial layer of FeSe films, especially after doping electrons. To unravel the underlying mechanism of interface-induced high temperature superconductivity, more theoretical studies are needed to elucidate the characteristics of spin fluctuations and their collaboration with electron-phonon coupling in this epitaxial system.

\begin{acknowledgments}

Project supported by the National Natural Science Foundation of China (Grant Nos.~11190024 and 11404383), the National Basic Research Program of China (Grant No. 2011CBA00112), the Fundamental Research Funds for the Central Universities, and the Research Funds of Renmin University of China (Grant No. 14XNLQ03).

\end{acknowledgments}

\end{document}